\documentclass[a4paper,11pt]{article}
\usepackage{pos}

\title{
Prospects for open heavy-flavour and quarkonium measurements with NA60+}
\ShortTitle{Open heavy-flavour and quarkonium with NA60+}

\author*[a]{Roberta Arnaldi for the NA60+ Collaboration}

\affiliation[a]{INFN, Sez. Torino,\\
 Via Pietro Giuria 1, I-10126, Torino, Italy}

\emailAdd{arnaldi@to.infn.it}

\abstract{
The high-intensity beams provided by the CERN SPS in a large range of energies offer a unique opportunity to investigate the region of the QCD phase diagram at high baryochemical potential.
The NA60+ experiment, proposed for taking data with heavy-ion collisions at the SPS in the next years, is in an ideal position to provide new insights into the QCD phase diagram, measuring rare probes via a Pb--Pb and p--A beam-energy scan, in the collisions energy interval $\sqrt{s_{NN}}$= 6-17~GeV.

NA60+ plans to measure the production of hidden and open charm hadrons and prospects on these measurements will be discussed.
Open charm hadrons will be measured from their decays into charged hadrons, reconstructed from the tracks in the silicon detectors of the vertex telescope.
This will enable high-precision measurements of the yield of D$^{0}$, D$^{+}$, and D$^{+}_{s}$ mesons, and of $\Lambda_{c}^{+}$ baryons, thus allowing us to constrain the transport properties of the QGP and the charm-quark hadronisation.
Charmonium states, J/$\psi$ and $\psi$(2S), will be measured through dimuon decays reconstructed with the muon spectrometer.
Hence, by measuring the charmonium yield in p--A and Pb--Pb collisions at different collision energies, NA60+ will have a unique opportunity to study the threshold energy for the onset of deconfinement.
}

\FullConference{HardProbes2023\\
 26-31 March 2023\\
 Aschaffenburg, Germany\\}

\begin{document}
\maketitle
\section{Introduction}
The study of the production of heavy-flavour (charm and beauty) hadrons both at RHIC and LHC are providing very interesting insights on the properties of the quark gluon plasma (QGP) at small baryochemical potential $\mu_{B}$. 
However, these studies would certainly benefit if extended such as to include new measurements in lower energy heavy-ion collisions. These new data would allow one to probe the medium at lower temperatures with respect to those obtained at LHC and RHIC colliders and to explore the region of finite $\mu_B$, in which the baryons of the colliding nuclei are stopped in the collision region. Very few results are available so far in this kinematic region. Concerning the heavy-flavour hadrons, there is only a first indirect open charm measurement by the NA60 Collaboration~\cite{NA60:2008dcb}, with an uncertainty of the order of 20$\%$, and an upper limit measurement on the $D^0$ production, by NA49~\cite{NA49:2005hwi}. Concerning quarkonium, many high precision results were reported by the NA50~\cite{NA50:2004sgj} and NA60~\cite{NA60:2006ncq} Collaborations, but only at top SPS energies, with the low energy range still to be investigated.

\section{The NA60+ experiment}
\begin{figure}
\begin{center}
\includegraphics[scale=0.6]{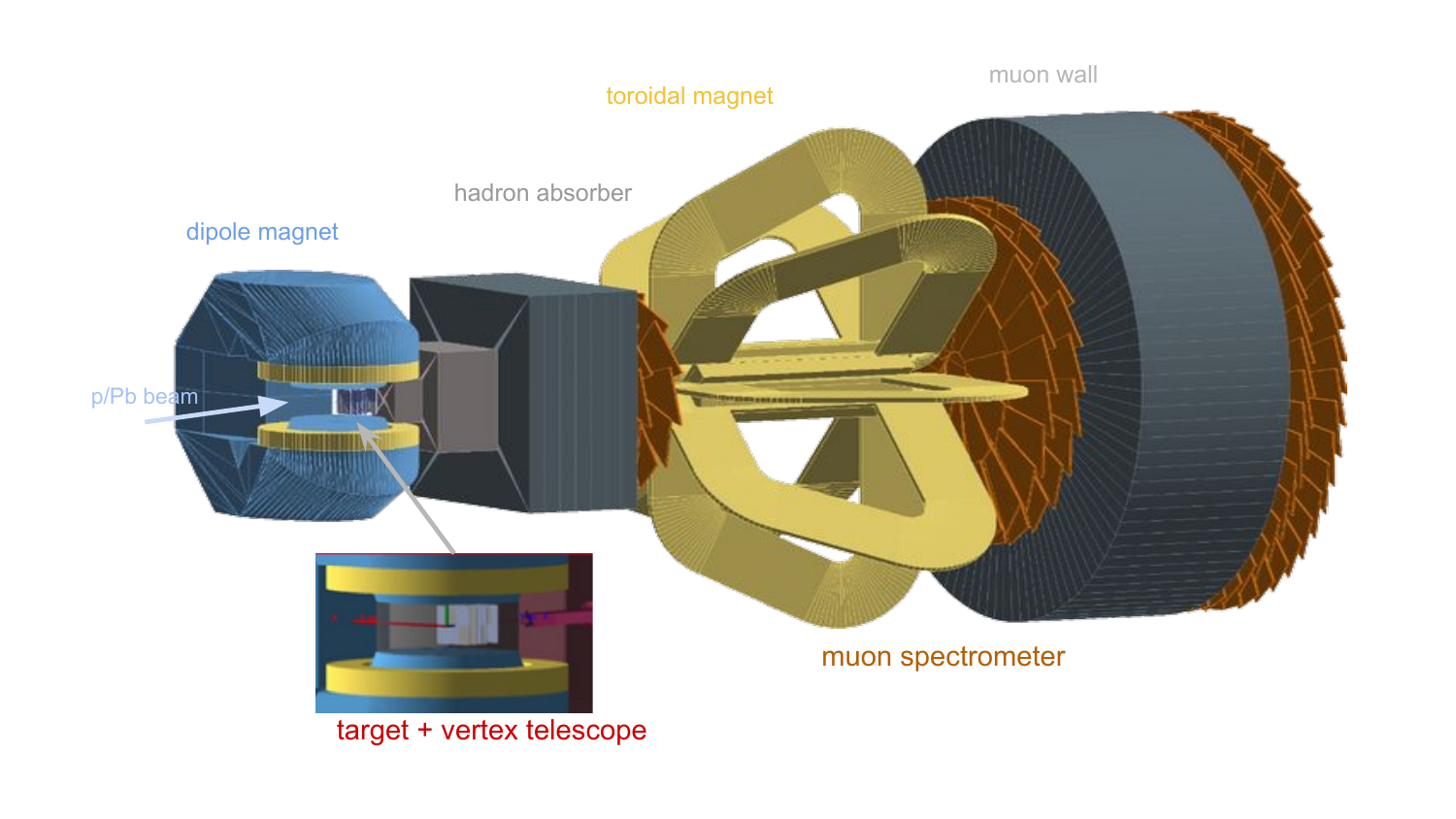}
\end{center}
\vskip -0.5cm
\caption{The NA60+ experimental setup. More details on the various detectors composing the apparatus can be found in \cite{NA60:2022sze}}
\label{fig:1}
\end{figure}

New high-precision measurements on heavy flavour hadrons will be within the reach of NA60+~\cite{NA60:2022sze}, a new experiment proposed at the CERN SPS. NA60+ will investigate proton-nucleus and Pb-Pb collisions, with high luminosity beams (i.e. $\sim10^{6}$ Pb/s) in the energy range E$_{\rm{lab}}$ = 20 - 158 AGeV, corresponding to the centre-of-mass energies $\sqrt{s_{NN}}$ = 6 - 17.3 GeV. The setup, shown in Fig.~\ref{fig:1}, will consist of a vertex spectrometer, made of silicon sensors, for the precise measurement and production angle of the produced charged particles, coupled with a muon spectrometer, to measure the muon tracks filtered by a thick hadron absorber, placed downstream of the vertex spectrometer. Details on the experimental apparatus can be found in \cite{NA60:2022sze}. 

The NA60+ experiment will, then, be able to investigate the dimuon spectrum from threshold to the charmonium mass region and to reconstruct the hadronic decays of charm and strange hadrons. 
It should be pointed out that the scope of the NA60+ experiment goes even beyond the study of open and closed heavy flavours, including also the study of electromagnetic probes of the QGP, via the measurement of the muon pair spectrum. More details can be found in ~\cite{ES_HP23}.

\section{Open heavy flavour measurements}
By investigating the production of open heavy flavours in nucleus-nucleus collisions at low collision energies, one can access the charm diffusion coefficient, extract information on charm thermalization and on the hadronization mechanism and eventually measure the total charm cross-section. The first parameter depends on the temperature of the medium and it is expected to be larger in the hadronic phase than in the QGP one, with a minimum around the pseudocritical temperature ($T_{\rm{pc}}\sim 155$ MeV). Hence, at the energies reached at the SPS accelerator, we should be able to explore the QGP properties closer to $T_{\rm {pc}}$. Furthermore, since at SPS energies the hadronic phase represents a larger fraction of the collision evolution with respect to the colliders case, the sensitivity to the hadronic interactions can be investigated. This measurement will also provide an input for precision estimates of the heavy-quark diffusion coefficient at collider energies. Studying the transverse momentum ($p_{T}$) distributions and the azimuthal anisotropy of D mesons, insights on the charm thermalization in a short-lived medium can also be addressed. Furthermore, the measurement of the relative abundances of the different charm-hadron species (i.e. comparing the production of strange and non-strange mesons or of baryons and mesons) can provide information on the hadronisation mechanisms. 
Finally, an accurate reconstruction of mesons and baryons ground states, such as the $D^0, D^+, D^{+}_{s}, \Lambda_c^+$ and possibly the $\Xi_c^{0,+}$, should also provide a precise measurement of the $c\bar{c}$ cross section, representing, therefore, a crucial reference for charmonium studies. It should be noted that, so far, the total charm production cross section has never been measured with high precision at top SPS energies or lower, because the yields at these energies are very small. 

Open-charm studies will be extremely relevant not only in Pb--Pb collisions, but also in proton-nucleus interactions, using nuclear targets with different atomic mass numbers, ranging from Be to Pb. These studies will open up the possibility to investigate the nuclear PDFs via D meson production. NA60+ will provide measurements in the the Bjorken-x range $0.1 < x_{Bj} <0.3$ at $Q^2\sim10-40$ GeV$^{2}$, i.e. in a kinematic range so far poorly constrained by data, dominated by anti-shadowing and EMC effects.  

\begin{figure}
\begin{center}
\includegraphics[scale=0.5]{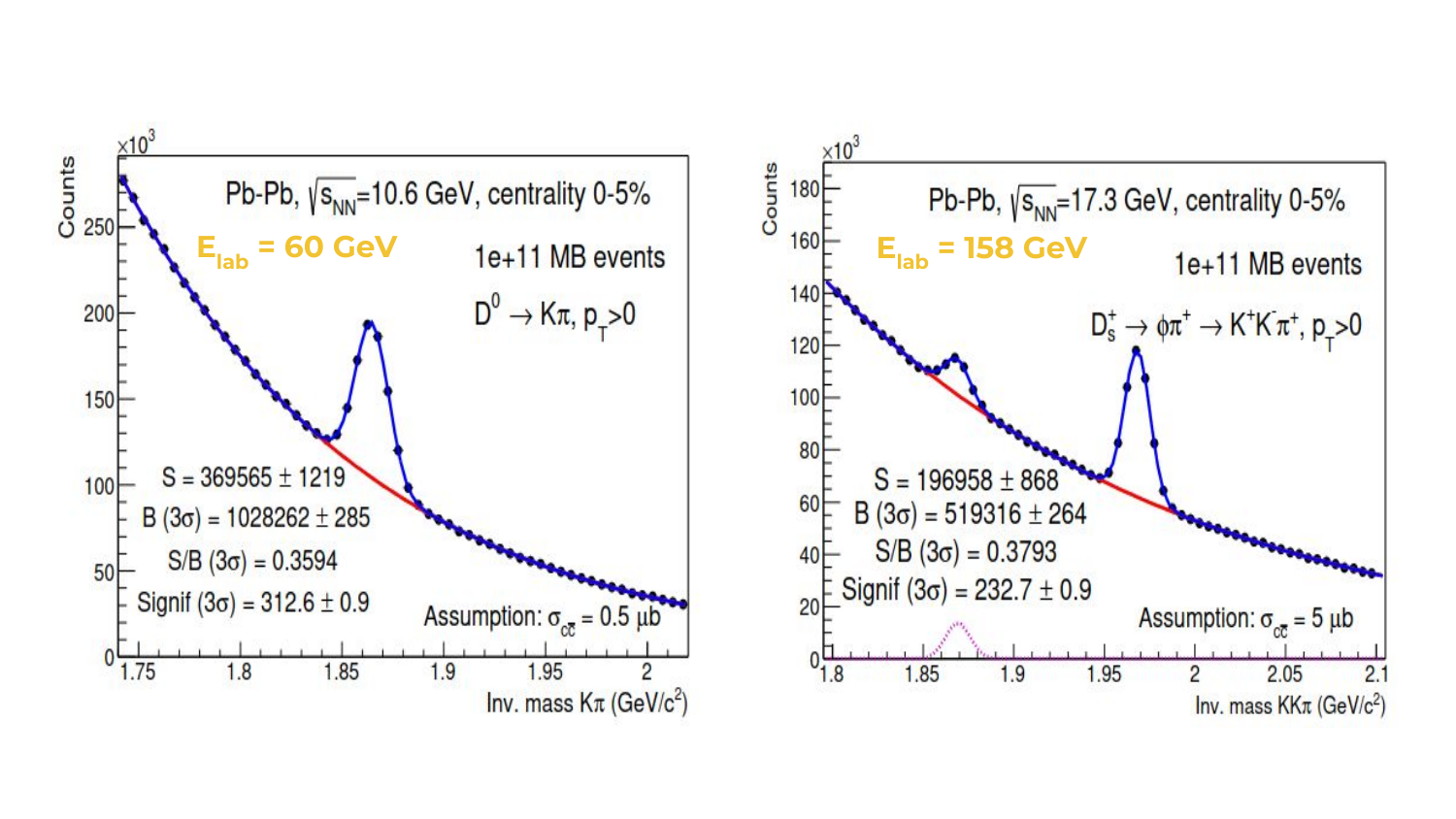}
\end{center}
\vskip -1cm
\caption{$D^{0}$ and $D^{+}_{s}$ invariant mass spectra, in central Pb-Pb collisions, at $E_{lab}$ = 60 GeV (left) and 158 GeV (right)}
\label{fig:2}
\end{figure}

NA60+ will measure the open-charm hadrons reconstructing, in the vertex telescope, their decays into two or three charged hadrons.  The huge combinatorial background, underlying these decays, can be reduced via geometrical selections on the displaced decay-vertex topology, exploiting the fact that the mean proper decay lengths $c\tau$ of open-charm hadrons are of the order of 60-310~$\mu$m. It should be noted that open-charm hadron studies are those which impose the strongest constraints on the design of the vertex telescope, since  a very good resolution on the track parameters is required. Performance studies show that in one month of data taking, corresponding to $\sim10^{11}$ minimum bias Pb-Pb collisions, it will be possible to measure the $D^0$ production through the decay channel $D^{0}\rightarrow K^{-}\pi^{+}$ with a statistical precision of the order of $1\%$ even at the lowest SPS energies. Similarly, the more challenging three-prongs decays will be investigated, such as the $D^{+}_{s} \rightarrow  \phi\pi^{+} \rightarrow K^{-}K^{+}\pi^{+}$, with a statistical precision of the order of few percent, also in this case. Example of invariant mass spectra are shown in Fig.~\ref{fig:2}. The $\Lambda_{c} ^{+}\rightarrow pK^{-}\pi^{+}$ measurement will also be accessible. The measure is quite challenging, given the short lifetime of the $\Lambda_c^+$ ($c\tau$ = 60 $\mu$m), and the inclusion, under study, of timing layers in the vertex telescope might help in the improvement of the NA60+ particle identification capabilities. 

\section{Quarkonium measurements}
The suppression of quarkonium states has been considered, since almost forty years, one of the key signatures of the QGP formation~\cite{Matsui:1986dk}. Experimental results obtained at top SPS energies confirm that the $J/\psi$ production is indeed suppressed and the observed $\sim$30$\%$ suppression is interpreted as due to the feed-down of the $\psi$(2S) and $\chi_{c}$ excited states. At the higher RHIC and LHC energies, a new mechanism, called regeneration, has been observed, leading to an enhancement of the quarkonium production, counterbalancing, at least partially, the suppression process. Quarkonium production is also affected by the cold nuclear matter effects, such as the nuclear modification of the PDFs (e.g. (anti-)shadowing) and the absorption of the (pre-)resonant state while crossing the nuclear matter. This latter mechanism is strongly energy-dependent and it is particularly sizable at SPS energies, because of the much larger crossing time, of the order of 0.5 fm/c at mid-rapidity, to be compared with $10^{-5}-10^{-2}$ fm/c at LHC energies, depending on the rapidity range. The precise determination of cold nuclear matter effects is crucial to understand the size of the hot matter effects at play and it is usually extensively studied via proton-nucleus collisions. 

The NA60+ experiment aims to study the J/$\psi$ production in Pb-Pb collisions, by performing an energy scan. Correlating the onset of the J/$\psi$ suppression with the temperature of the system, as measured via the thermal dimuons (this is one of the main topics that will be investigated by the NA60+ experiment~\cite{ES_HP23}), the threshold temperature for the melting of the resonance can be accessed. Proton-nucleus collisions, collected at the same collision energy as the A-A data, will allow a detailed study of the cold nuclear matter effects, whose contribution is expected to increase, as already mentioned, when the collision energy is decreased. It will be important not only to disentangle the various cold nuclear matter effects at play, but also to define a precise reference to understand the role of hot matter effects on charmonium. Finally, the study of the J/$\psi$ production at SPS energies will also be a testing ground for the observation of the intrinsic charm component in the nucleon wavefunction. This effect should lead to an enhanced charm production at large $x_{F}$. While at collider energy the region where the intrinsic charm should be observed is at very forward rapidity, in fixed target experiments, at SPS energies, it should be closer to mid-rapidity, and hence more easily detectable. Calculations~\cite{Vogt:2021vsc} indicate that already with a probability of intrinsic charm contribution in the proton of 0.1\%, the effect of the intrinsic charm should be dominant.

\begin{figure}
\begin{center}
\includegraphics[scale=0.55]{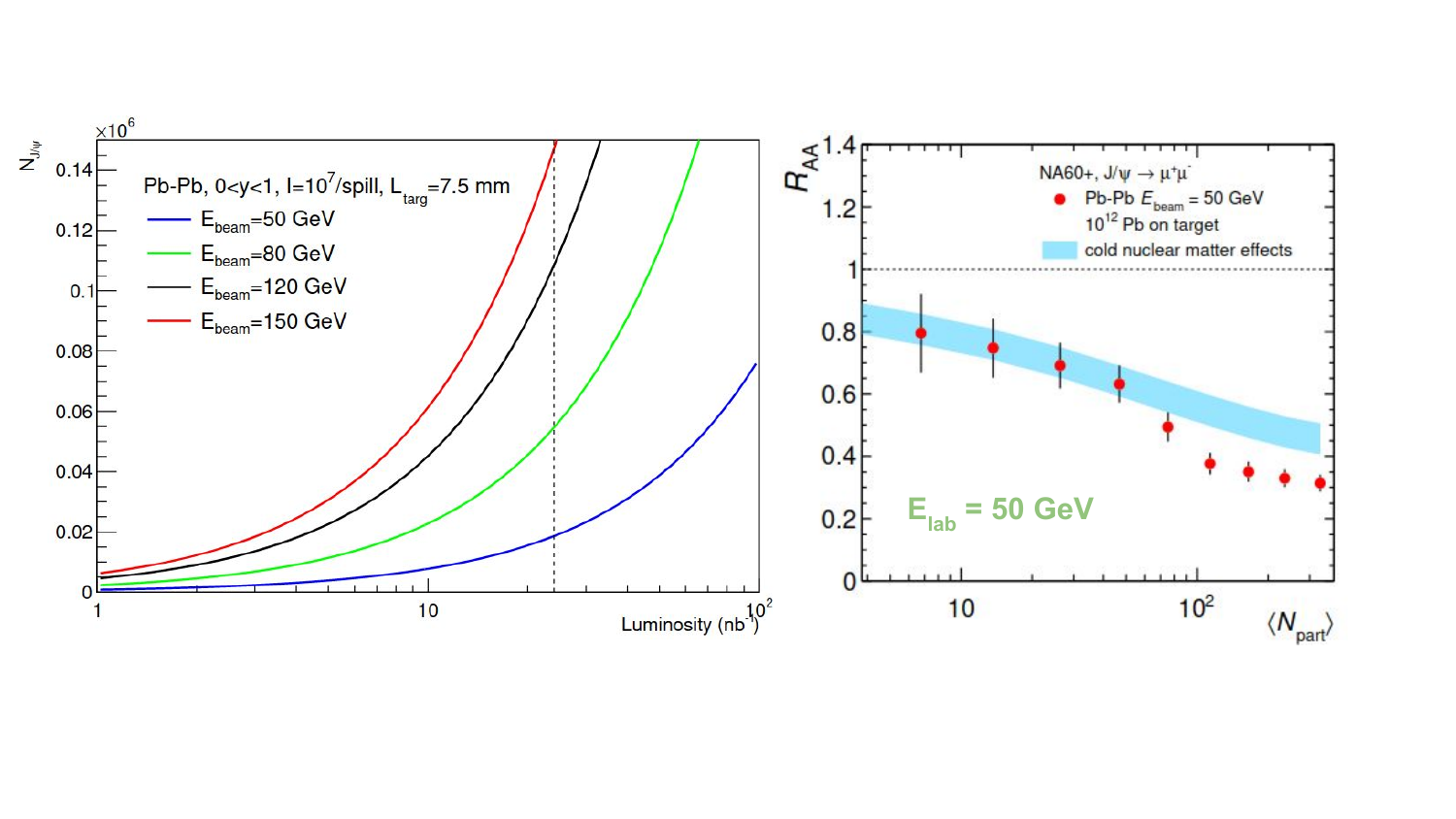}
\end{center}
\vskip -2cm
\caption{Number of expected J/$\psi$ collected under the assumptions described in the text, as a function of the luminosity (left); performance study of the J/$\psi$ nuclear modification factor $R_{AA}$ as a function of centrality, at $E_{\rm{lab}}$ = 50 GeV. The light blue band corresponds to the cold nuclear matter effects, as obtained from a performance study based on realistic expectations for the p-A data taking.}
\label{fig:3}
\end{figure}

J/$\psi$ and $\psi$(2S) will be measured in NA60+ through their decay in a pair of muons, reconstructed by matching the tracks in the muon spectrometer with those reconstructed in the vertex telescope. The expected mass resolution will be of the order of 30 MeV/c. The $\chi_c$ will be measured in the decay channel $\chi_c \rightarrow J/\psi \gamma$, with the $\gamma$ measured via its conversion into a lepton pair in the vertex telescope.
To cope with the low production cross sections at low $\sqrt{s}$, a high luminosity is required. As shown in Fig.~\ref{fig:3} (left), performance studies based on one month of data taking, with  $I_{\rm{beam}} = 10^7$ Pb/spill, and a total luminosity $L_{\rm{int}}$ = 24  $nb^{-1}$, assuming a factor 3 overall suppression for quarkonium, as educated guess, we will collect $10^4 - 10^5$ J/$\psi$, depending on the collision energy. This statistics will allow us to perform an accurate study of the J/$\psi$ suppression from top SPS energies, down to the lowest available ones. As shown in Fig.~\ref{fig:3} (right), we should be able to determine the onset for the suppression even at low $\sqrt{s_{NN}}$.

A similar study on the $\psi$(2S) will also be within reach, at least down to $E_{\rm{lab}}$ = 100 GeV. Reaching precise results also at lower beam energies would require larger beam intensities or longer data taking periods. Studies for the detection of the $\chi_c \rightarrow \mu\mu\gamma$ decays are ongoing and they'll enrich the NA60+ charmonium program.

\section{Conclusions}
Open charm and quarkonium physics has proven to be of extreme interest in the understanding the properties of the QGP. However, while plenty of results are available at the RHIC and LHC collider energies, we are so far lacking results below top SPS energies. Hence, measurements carried on between $6 < \sqrt{s_{NN}} < 17$ GeV/c will be mandatory to expand our understanding towards the region of high-$\mu_B$. The NA60+ experiment, proposed at CERN SPS starting from 2029, after the LHC Long Shutdown 3, is certainly well positioned to perform this kind of studies and will certainly help to shed further light on the heavy-flavour physics, complementing the understanding so far achieved after a few decades of RHIC and LHC results.

\end{document}